# CORE – A New Method for Solving Hamiltonian Lattice Systems

Colin J. Morningstar[†] and Marvin Weinstein[‡]

[†] Department of Physics & Astronomy, University of Edinburgh, Edinburgh EH9 3JZ, Scotland
[‡] Stanford Linear Accelerator Center, Stanford University, Stanford, California 94309

**Abstract**

The COntractor REnormalization group (CORE) approximation, a new method for solving Hamiltonian lattice systems, is introduced. The approach combines variational and contraction techniques with the real-space renormalization group approach and is systematically improvable. Since it applies to lattice systems of infinite extent, the method is suitable for studying critical phenomena and phase structure; systems with dynamical fermions can also be treated. The method is tested using the 1+1-dimensional Ising model.

## 1. Introduction

Perturbative methods are inadequate for investigating many current problems in high energy and condensed matter physics, such as the confinement of quarks and gluons in QCD. Thus, the development of new nonperturbative tools is important. The COntractor REnormalization group (CORE) approximation, a new nonperturbative method for studying Hamiltonian lattice systems, is presented in this talk. The method is a hybrid of contraction, variational, cluster, mean-field, and block renormalization-group techniques. It is systematically improvable and applies to lattice systems of infinite extent, enabling direct study of phase structure and critical phenomena. Dynamical fermions can be treated without problem.

We briefly describe the method, then apply two variants of the CORE approximation to the 1+1-dimensional Ising model.

## 2. Description of the Method

The success of any variational calculation, especially one involving an infinite number of degrees of freedom, depends crucially on choosing a good trial state. An algorithm for building trial states suitable for lattice systems is the Hamiltonian real-space renormalization group (RSRG) method [1]. In this approach, the lattice is partitioned into blocks including a few sites and the block-Hamiltonians are diagonalized. The Hilbert space is then *thinned* by discarding all high-energy states, retaining only those states which can be constructed from tensor products of some small subset of low-lying block eigenstates, and an effective Hamiltonian which describes the mixing of the remaining states is computed. This thinning process is repeated again and again until the effective Hamiltonian takes a fixed form which can be diagonalized.

Unfortunately, simple RSRG truncation procedures often have difficulties accurately describing the long-wavelength modes on the full lattice because they badly underestimate the block-to-block mixings. Past approaches to overcoming this problem have concentrated on using larger blocks, increasing the number of states retained per block, or introducing more sophisticated truncation schemes. The CORE approximation is a new approach to this problem which emphasizes simplicity and versatility; it frees one from the need to develop clever truncation schemes and allows the use of manifestly gauge-invariant RSRG schemes when studying lattice gauge theories.

The basic idea of the CORE approach is to steer the RSRG iteration using contraction techniques. An important part of this steering process is reliably approximating the expectation value

$$\mathcal{E}(t) = \frac{\langle \Phi_{\text{var}} | e^{-tH} H e^{-tH} | \Phi_{\text{var}} \rangle}{\langle \Phi_{\text{var}} | e^{-2tH} | \Phi_{\text{var}} \rangle}, \qquad (1)$$

which tends to the lowest eigenvalue $\epsilon_0$ of Hamiltonian $H$ as $t \to \infty$, assuming the trial state $|\Phi_{\text{var}}\rangle$ has non-vanishing overlap with the ground state of $H$. Since $e^{-tH}$ cannot be computed exactly, it is replaced in the CORE method by an operator $T(t)$ which closely approximates $e^{-tH}$ for $t$ in some range $0 < t < t_{max}$ and which can be easily evaluated. A procedure for constructing such an operator has been described in Ref. [2]. In this procedure, one expresses $e^{-tH}$ as a symmetric product of explicitly computable terms; for example, if $H = H_1 + H_2$, where $e^{-tH_1}$ and $e^{-tH_2}$ can be evaluated exactly, then

$$e^{-tH} = e^{-tH_1/2} e^{-tH_2/2} e^{C_3(t)} e^{-tH_2/2} e^{-tH_1/2}, \qquad (2)$$

where $C_3(t)$ is order $t^3$ or higher. To construct $T(t)$, one then either replaces $e^{C_3(t)}$ by the identity operator or retains low-order terms in $C_3(t)$, rewriting their exponential again as a symmetric product of computable terms. A given contractor $T(t)$ can also be improved by using $T_p(t) = [T(t/p)]^p$.

Having chosen a contractor $T(t)$, a variational best estimate for $\epsilon_0$ can be obtained by minimizing

$$\mathcal{E}_T(t) = \frac{\langle \Phi_{\text{var}} | T(t) H T(t) | \Phi_{\text{var}} \rangle}{\langle \Phi_{\text{var}} | T(t)^2 | \Phi_{\text{var}} \rangle} \qquad (3)$$

with respect to $t$ and any parameters in $|\Phi_{\text{var}}\rangle$. For a trial state $|\Phi_{\text{var}}\rangle = \sum_{j=1}^n \alpha_j |\phi_j\rangle$, where $\{|\phi_j\rangle\}$ is any set of orthonormal states, minimizing $\mathcal{E}_T(t)$ with respect to the $\alpha_j$ parameters is equivalent to solving the generalized eigenvalue problem

$$\det\left(\llbracket T(t) H T(t) \rrbracket - \lambda \llbracket T(t)^2 \rrbracket\right) = 0, \qquad (4)$$

where $\llbracket \ldots \rrbracket$ denotes truncation to the subspace spanned by the $|\phi_j\rangle$ states. Hence, we can replace the problem of finding the best trial state by that of diagonalizing the *effective Hamiltonian*

$$H_{\text{eff}}(t) = \llbracket T(t)^2 \rrbracket^{-1/2} \llbracket T(t) H T(t) \rrbracket \llbracket T(t)^2 \rrbracket^{-1/2}. \qquad (5)$$

Developing *this* operator in the RSRG iteration instead of $\llbracket H \rrbracket$ is the key innovation of the CORE approach.

The effective Hamiltonian cannot be exactly determined. The last step in the CORE approach is to apply *cluster* techniques to approximate $H_{\text{eff}}(t)$ (see Ref. [3] and references cited therein). Essentially, this involves evaluating $H_{\text{eff}}(t)$ on increasingly-larger, connected sub-lattices and using the principle of inclusion-exclusion to appropriately combine the results for the full lattice.

In summary, CORE is an iterative blocking and thinning process, developing the low-lying physics in a sequence of effective sub-Hamiltonians $H_{\text{eff}}^{(n)}(t_n^*)$ using the recursion relation

$$H_{\text{eff}}^{(n+1)}(t) = R_n(t) \llbracket T^{(n)}(t) H_{\text{eff}}^{(n)}(t_n^*) T^{(n)}(t) \rrbracket R_n(t), \qquad (6)$$

where $R_n(t) = \llbracket T^{(n)}(t)^2 \rrbracket^{-1/2}$, the contractor $T^{(n)}(t)$ approximates $\exp[-tH_{\text{eff}}^{(n)}(t_n^*)]$, and $t_n^*$ is a best value for $t$ selected for each RG iteration in some manner: minimizing $H_{\text{eff}}^{(n)}(t)$ in a simple product state is one possibility. As the recursion proceeds, the effective Hamiltonian evolves eventually into a simple form which can be easily diagonalized, yielding estimates of the ground state energy and the energies of some low-lying excited states.

CORE can also be used to estimate the vacuum expectation value of an extensive operator $O$. Using the same RSRG transformations as for $H$, one first computes the sequence of effective operators $O_{\text{eff}}^{(n)}(t_n^*)$. Once $H_{\text{eff}}$ has evolved sufficiently such that its ground state can be found, the matrix element of $O_{\text{eff}}$ in the ground state of $H_{\text{eff}}$ then yields the desired expectation value.

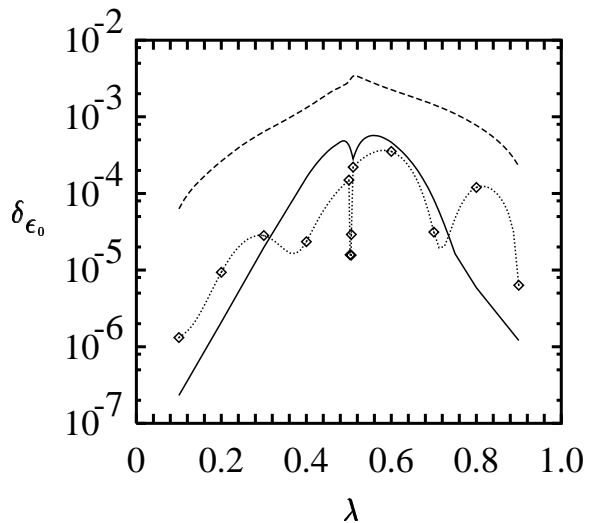

**Figure 1.** Fractional error $\delta_{\epsilon_0}$ in the ground-state energy density estimates against $\lambda$. Results using $T_1^2$ (dashed curve), $T_1^{16}$ (solid), and $T_2^{12}$ (diamonds with dotted curve) are shown.

## 3. The 1+1-Dimensional Ising Model

The Ising model in $1 + 1$ dimensions is often used as a testing ground for new calculational methods. Its Hamiltonian is given by

$$H_{\text{Ising}} = -\sum_j \left[ c_\lambda \sigma_z(j) + s_\lambda \sigma_x(j) \sigma_x(j+1) \right], \qquad (7)$$

where $j$ labels the sites in the infinite chain, $c_\lambda = \cos(\lambda\pi/2)$, and $s_\lambda = \sin(\lambda\pi/2)$, for $0 \leq \lambda \leq 1$. A second-order phase transition occurs in this model at $\lambda = 1/2$.

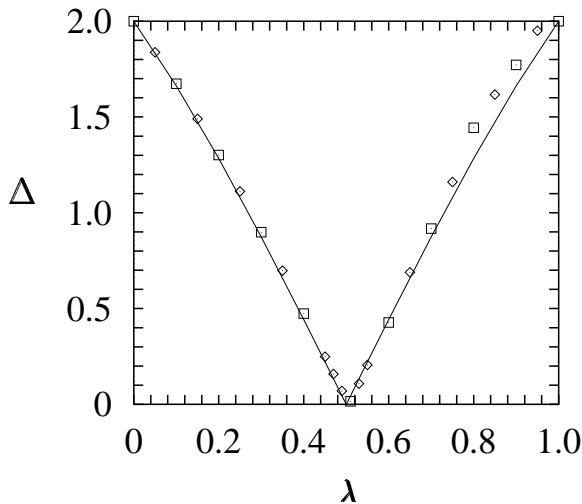

Figure 2. Mass gap estimates $\Delta$ against $\lambda$. The diamonds and squares indicate CORE estimates obtained using $T_1^{16}$ and $T_2^{12}$, respectively. The exact mass gap appears as a solid curve.

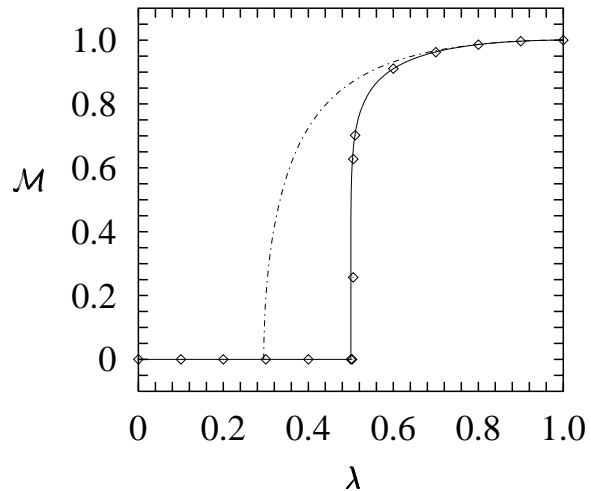

Figure 3. Magnetization $\mathcal{M}$ against $\lambda$. The diamonds indicate CORE estimates obtained using $T_2^{12}$, the solid curve shows the exact magnetization, and the dot-dashed curve shows the estimates from mean-field theory.

When $\lambda < 1/2$, the order parameter $\langle \sigma_x(j) \rangle = 0$ and the ground state is unique. For $\lambda > 1/2$, spontaneous symmetry breaking occurs and the order parameter takes values $\langle \sigma_x(j) \rangle = \pm(1 - \cot^2(\lambda\pi/2))^{1/8}$.

We tested the CORE approximation in two different applications to the Ising model. In both applications, the Hilbert space was thinned to the lowest two eigenstates in each block, the cluster expansion of $H_{\text{eff}}(t)$ was truncated after three-block clusters, and $t$ was fixed by minimizing the expectation value of $H_{\text{eff}}$ in a mean-field state. Two-site blocking was used in the first application, and blocks containing three sites were used in the second application. The contractor for the first application was $T_1(t) = S_1^\dagger(t) S_1(t)$, with $S_1(t) = \prod_\alpha \{\prod_j [1 + \tanh(c_\alpha t/2) O_\alpha(j)]\}$, where $c_\alpha$ are couplings, $\alpha$ labels the different types of operators $O_\alpha(j)$, such as $\sigma_z(j)$ and $\sigma_x(j)\sigma_x(j+1)$, and $j$ is a site label. The second contractor used was $T_2(t) = S_2^\dagger(t) S_2(t)$ with $S_2(t) = \exp(-tV/2) \exp(-tH_b/2)$, where $H_b$ contains all intra-block interactions and $V$ contains all inter-block operators (those which cross block boundaries). Note that $\exp(-tH_b/2) = \prod_p \exp(-tH_b(p)/2)$ and $\exp(-tV/2) = \prod_p \exp(-tV(p)/2)$, where $p$ labels the blocks. Calculations were done using $T_1^n(t/n)$ and $T_2^n(t/n)$ for various values of $n$.

Fractional errors $\delta_{\epsilon_0} = |(E_0 - \epsilon_0)/\epsilon_0|$ in the ground-state energy estimates $E_0$ from both variants of the CORE approach are shown in Fig. 1. Selected estimates for the mass gap $\Delta$ and magnetization $\mathcal{M} = |\langle \sigma_x(j) \rangle|$, for some site $j$, are compared to the exactly-known results in Figs. 2 and 3. Considering that only the first three terms in the cluster expansion are included in the calculations, the accuracy of the results is striking. The CORE approximation reproduces the correct location of the critical point with remarkable precision. Including more terms in the cluster expansion should significantly improve these results.

## 4. Conclusion

We believe that the CORE approximation will prove to be a powerful tool for studying nonperturbative systems. An exciting feature of the method is that it can be used to analyze systems containing dynamical fermions, systems which resist treatment by present stochastic means. We are presently extending the method for use with lattice field theories.

## 5. Acknowledgments

This work was supported by the NSERC of Canada, the U. S. DOE, Contract No. DE-AC03-76SF00515, and the UK SERC, grant GR/J 21347.